\documentclass[aps,prl,twocolumn,showpacs,floatfix,preprintnumbers,amsfont,amsmath,amssymb,nofootinbib, superscriptaddress]{revtex4}

\usepackage{color}
\usepackage[font=small]{caption,subfig}
\usepackage{hyperref}
\usepackage[normalem]{ulem}
\usepackage{lipsum}
\usepackage{url}
\usepackage{float}
\usepackage{ctable}
\usepackage{graphicx}
\usepackage{bm}
\usepackage[font=small,
   justification=justified,
   format=plain]{caption} % 'format=plain' avoids hanging indentation

\newcommand{\sk}[1]{}

\newcommand{\refeq}[1]{Eq.~(\ref{eq:#1})}          
          
\newcommand{\reffig}[1]{Fig.~\ref{fig:#1}}

\def\VEV#1{\left\langle #1 \right\rangle}

\newcommand{\be}{\begin{equation}}
\newcommand{\ee}{\end{equation}}
\newcommand{\ba}{\begin{eqnarray}}
\newcommand{\ea}{\end{eqnarray}}
\newcommand{\en}{\nonumber\\}

\newcommand{\bfn}{{\boldsymbol{n}}}

\newcommand{\aap}{Astron. Astrophys.}

\allowdisplaybreaks % let equations span multiple pages

\begin{document}

\title{On the waveforms of gravitationally lensed gravitational waves}

\author{Liang Dai}
\thanks{NASA Einstein Fellow}
\affiliation{\mbox{School of Natural Sciences, Institute for Advanced Study, 1 Einstein Drive, Princeton, New Jersey 08540, USA}}
\author{Tejaswi Venumadhav} 
\affiliation{\mbox{School of Natural Sciences, Institute for Advanced Study, 1 Einstein Drive, Princeton, New Jersey 08540, USA}}

\date{\today}

%%%%%%%%%%%%%%%%%%%%%%%%%%%%%%%%%%%%%%%%%%%%%%%%%%%%%%%%%%%%%%%%%%%%%%%%%%%%%%%

\begin{abstract}

Strong lensing by intervening galaxies can produce multiple images of gravitational waves from sources at cosmological distances. These images acquire additional phase-shifts as the over-focused wavefront passes through itself along the line of sight. Time-domain waveforms of Type-II images (associated with saddle points of the time delay) exhibit a non-trivial distortion from the unlensed waveforms. This phenomenon is in addition to the usual frequency-independent magnification, and happens even in the geometric limit where the wavelength is much shorter than the deflector's gravitational length scale. Similarly, Type-III images preserve the original waveform's shape but exhibit a sign flip. We show that for non-precessing binaries undergoing circular inspiral and merger, these distortions are equivalent to rotating the line of sight about the normal to the orbital plane by $45^\circ$ (Type II) and $90^\circ$ (Type III). This effect will enable us to distinguish between the different topological types among a set of multiple images, and give us valuable insight into the lens model. Furthermore, we show that for eccentric binaries, the waveform of a Type-II image is distorted in a manner that is inequivalent to a change of the source's orbital parameters.

\end{abstract}

%%%%%%%%%%%%%%%%%%%%%%%%%%%%%%%%%%%%%%%%%%%%%%%%%%%%%%%%%%%%%%%%%%%%%%%%%%%%%%%

\maketitle

%%%%%%%%%%%%%%%%%%%%%%%%%%%%%%%%%%%%%%%%%%%%%%%%%%%%%%%%%%%%%%%%%%%%%%%%%%%%%%%

{\it Introduction.}---Gravitational waves (GW) are a promising new observational probe of the dynamics of compact stellar objects. The recent detections by LIGO~\cite{Abbott:2016blz, Abbott:2016nmj, 2016PhRvX...6d1015A} hint at a significant population of stellar-mass black hole mergers, which forthcoming advanced ground-based detectors will see out to cosmological distances. In addition, proposed space-based observatories would be sensitive to GWs from distant supermassive black holes, emitted either during their mutual coalescence, or their capture of smaller compact objects \cite{2017arXiv170200786A}.

Sources at cosmological distances can be strongly lensed by collapsed structures  along the line of sight. In general, such strongly lensed sources are (de)magnified and multiply imaged. The overall optical depth to strong lensing for cosmological sources is dominated by galactic-scale ($M \gtrsim 10^{10} M_\odot$) halos~\cite{2003ApJ...595..603L}. For astrophysical sources, lensing by these halos is well described by geometrical optics (except at the low frequency end of the band for space-based detectors~\cite{Takahashi:2003ix}), which approximates the propagation of GWs in terms of families of rays that are normal to wavefronts.

Previous works on lensing (including ours \cite{Dai:2016igl}) have implicitly assumed that in the time-domain, strain waveforms of lensed images are just the unlensed ones multiplied by overall factors of the square root of the usual magnification factors for electromagnetic flux. In this {\it Letter}, we explore the observational consequences of an important exception where (even in the short-wavelength limit, in which geometrical optics works) lensed gravitational waveforms appear inequivalent to the unlensed ones. This effect originates in the deformation of wavefronts as they pass through caustics, which causes an additional topological phase shift in the waveforms \cite{2012PhRvD..86f4030Z}.

When the lensing systems are galactic-mass halos, multiple images are separated by time delays of typically a few days to months. Owing to the small amount of time that mergers spend `in band' for detectors, these images would be triggered as independent events. Without additional knowledge of the source redshift, the magnification of each individual image is degenerate with the intrinsic mass scale and redshift of the source~\cite{Dai:2016igl}. In such cases, strong lensing would have to be inferred from the coincidence of positions on the sky and the detailed shapes of the signals. The effect described in this paper is important for the latter consideration, since GW observations directly access the waveforms (unlike electromagnetic observations at higher frequencies which typically return intensities).

\begin{figure*}[t]
  \subfloat[]{
    \includegraphics[width=0.8\columnwidth]{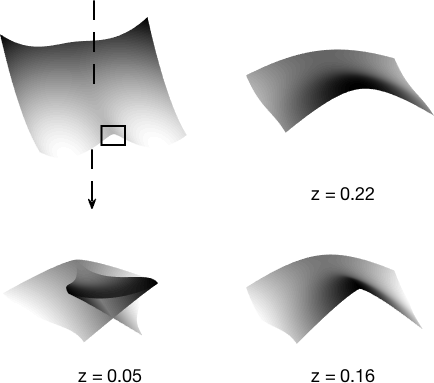}
    \label{fig:wfpinch}
   } 
   ~
   \subfloat[]{
     \includegraphics[width=0.8\columnwidth]{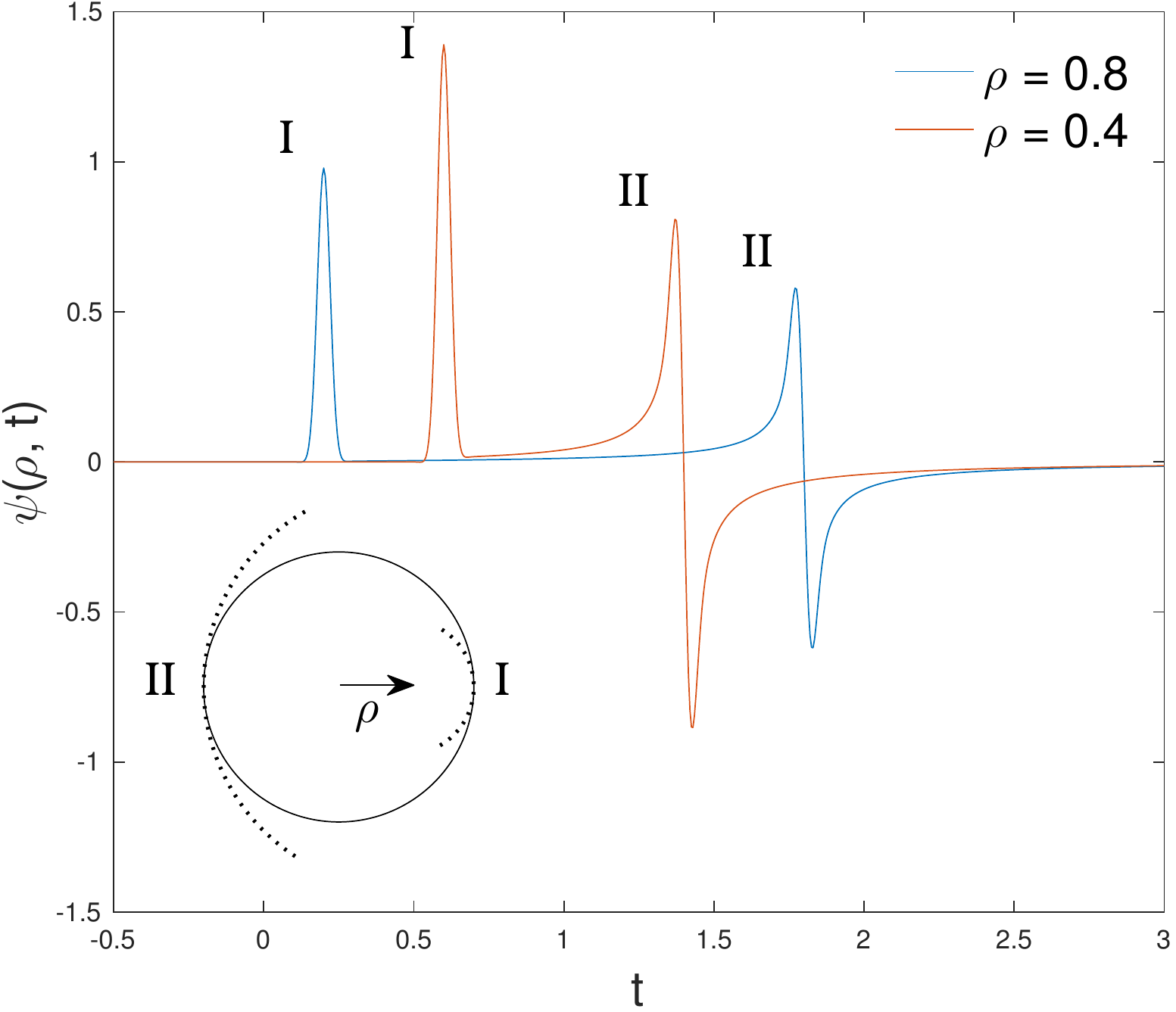}
    \label{fig:cylinechoes}     
   }
\caption{\label{fig:wfechoes} {\em (a) Top left sub-panel:} Downward propagating wavefront after a point source at $z_s = 1$ is lensed by a cored elliptical isothermal lens located at $z_d = 0.25$. {\em Clockwise starting from top right:} Zoomed in views into the marked section at successively lower redshifts. The wavefront pinches and intersects itself at $z=0.16$, and at lower redshifts, is `creased' at critical curves. {\em (b):} Amplitude of a cylindrically focused wave as a function of time at two observing locations of different radii. Inset shows the wavefront geometry---solid line is the incoming wavefront at time $t = 0$. Roman indices show the type of the image. As the dotted arcs centered on the observing location show, the Type-II image is a local maximum of the arrival time along the $\hat{\bm \phi}$ direction, and has a different waveform from the Type-I image.}
\end{figure*}

%%%%%%%%%%%%%%%%%%%%%%%%%%%%%%%%%%%%%%%%%%%%%%%%%%%%%%%%%%%%%%%%%%%%%%%%%%%%%%%

{\it Image waveforms.}---\sk{Analogous to the case of electromagnetic waves,}The lensing of GWs can be understood in terms of the Kirchhoff diffraction integral for radiation originating from a given source and reaching a given observer~\cite{PhysRevD.59.083001}. In the geometric limit, the diffraction integral picks up contributions from the vicinity of extremal points of the Fermat potential, or the time delay, over a family of trajectories~\cite{1986ApJ...310..568B}. In the vicinity of these extremal trajectories, the diffraction integral is a two-dimensional Gaussian integral, and the associated image is of Type I, II, or III depending on whether the trajectory is a local mimimum, saddle point, or maximum of the time delay. For the $j^{\rm th}$ image, the complex strain amplitude is the source pulse convolved with~\cite{schneider1992gravitational}
\ba
F_j(f) & = & \left| \mu_j \right|^{1/2}\,\exp\left[ i\,2\,\pi\,f\,T_j - i\,\pi\,n_j\,{\rm sgn}(f) \right].
\ea 
Here $f$ is the observed wave frequency, $T_j$ is the total comoving travel time along the null trajectory for the $j^{\rm th}$ image, $\mu_j$ is the signed magnification factor as given by the inverse of the determinant of the lensing Jacobian matrix. The extra phase shift $\pi\,n_j\,{\rm sgn}(f)$ arises from the complex Gaussian integral---$n_j$ is called the Morse index~\cite{morse1934calculus,ambrose1961index}, and equals half the number of negative eigenvalues of the lensing Jacobian matrix, i.e., $n_j = 0$ for minima, $1$ for maxima, and $1/2$ for saddle points. 

For a Type-I image, there is no extra phase shift, and the lensed waveform is equivalent to the unlensed one apart from the $2\,\pi\,f\,T_j$ term, which accounts for the travel time to the observer (the magnification factor $\mu_j$ does not depend on $f$). For a Type-III image, the phase shift equals $\pi$ for all frequencies, and thus the waveform flips its sign. This flip is measurable by a GW detector, which is sensitive to amplitudes and not just intensities. A non-trivial effect occurs for a Type-II image, for which all positive-frequency components are shifted by a phase $-\pi/2$, and all negative-frequency components are shifted by $\pi/2$. This effect differs from that of an overall phase factor due to the opposing shifts of the positive and negative frequencies, and distorts the intrinsic waveform, $h(t)$, to its Hilbert transform, i.e.,
\ba
\tilde h(t) & = & - \int^{+\infty}_{-\infty}\,df\,{\rm sgn}(f)\,i\,e^{-i\,2\,\pi\,f\,t} h(f),
\ea
where $h(f) = \int^{+\infty}_{-\infty}\,h(t)\,e^{i\,2\,\pi\,f\,t}$ is the Fourier transformation of the intrinsic waveform (for one polarization), and the lensed waveform $\tilde h(t)$ is real in the time domain. \sk{Note that for a superposition of a narrow range of wave frequencies, the distorted wave $\tilde h(t)$ resembles but is not equivalent to the time derivative of the original wave $h(t)$, which amounts to multiplying $- i\,2\,\pi\,f$ in Fourier space.}

To understand the extra phase factor, let us consider the evolution of a scalar wave's wavefront. Figure \ref{fig:wfpinch} shows a downward propagating wavefront after a point source at $z_s = 1$ is lensed by a cored elliptical isothermal lens (\cite{1994A&A...284..285K}; axis-ratio $q = 0.2$, velocity dispersion $\sigma_v = 300 \ {\rm km/s}$, and core-radius $b = 1.1 \ {\rm kpc}$) located at $z_d = 0.25$. The wavefront intersects itself along a line, and subsequently passes a given point multiple times \cite{1983A&A...128..156K}. If we approximate the wavefront as flat in the second direction at the moment of intersection, it locally evolves as a converging cylindrical front. Figure \ref{fig:cylinechoes} shows the amplitude of a converging cylindrical scalar wave through its focusing (the case with exact symmetry can be solved analytically). We see that the time-domain waveform of the second image differs from that of the first.

%%%%%%%%%%%%%%%%%%%%%%%%%%%%%%%%%%%%%%%%%%%%%%%%%%%%%%%%%%%%%%%%%%%%%%%%%%%%%%%

{\it Lensing in parameter space.}---We now study how waveforms of Type-II images of GW bursts differ from the unlensed ones in parameter space. For any lens model, the image that arrives the earliest is always a Type-I image. Hence observationally, the extra topological phase-shifts we quote can be considered as relative to this image.

A template search using the unlensed waveform does not yield an ideal match. To see this, consider the matched-filtering signal-to-noise (SNR) at lag $\tau$, ${\rm SNR}(\tau) = \VEV{ \tilde h | h}_\tau / \sqrt{\VEV{h | h}}_{\tau = 0}$ (the overlap between data $a$ and template $b$ at lag $\tau$ is $\VEV{a | b}_\tau \equiv 2\,\int^{+\infty}_{-\infty}\,df\,a(f)\,b^*(f)\,e^{i\,2\,\pi\,f\,\tau}\,[S_n(f)]^{-1}$, where $S_n(f) = S_n(-f)$ is the detector's one-sided noise power spectrum). At zero lag, i.e., $\tau = 0$, the SNR vanishes, since $\VEV{\tilde h| h}_{\tau = 0} = - 2\,i\, \int^{+\infty}_{-\infty}\,df\,{\rm sgn}(f)\,|h(f)|^2\,[S_n(f)]^{-1} = 0$. A better SNR is achieved for $\tau\neq 0$, but it is reduced compared to the unlensed case. We will see that the Type-II image's waveform instead resembles the unlensed waveform with different source parameters.

%%%%%%%%%%%%%%%%%%%%
\begin{figure}[t]
  \begin{center}
   % \hspace{-1.0cm}
   \includegraphics[width=\columnwidth]{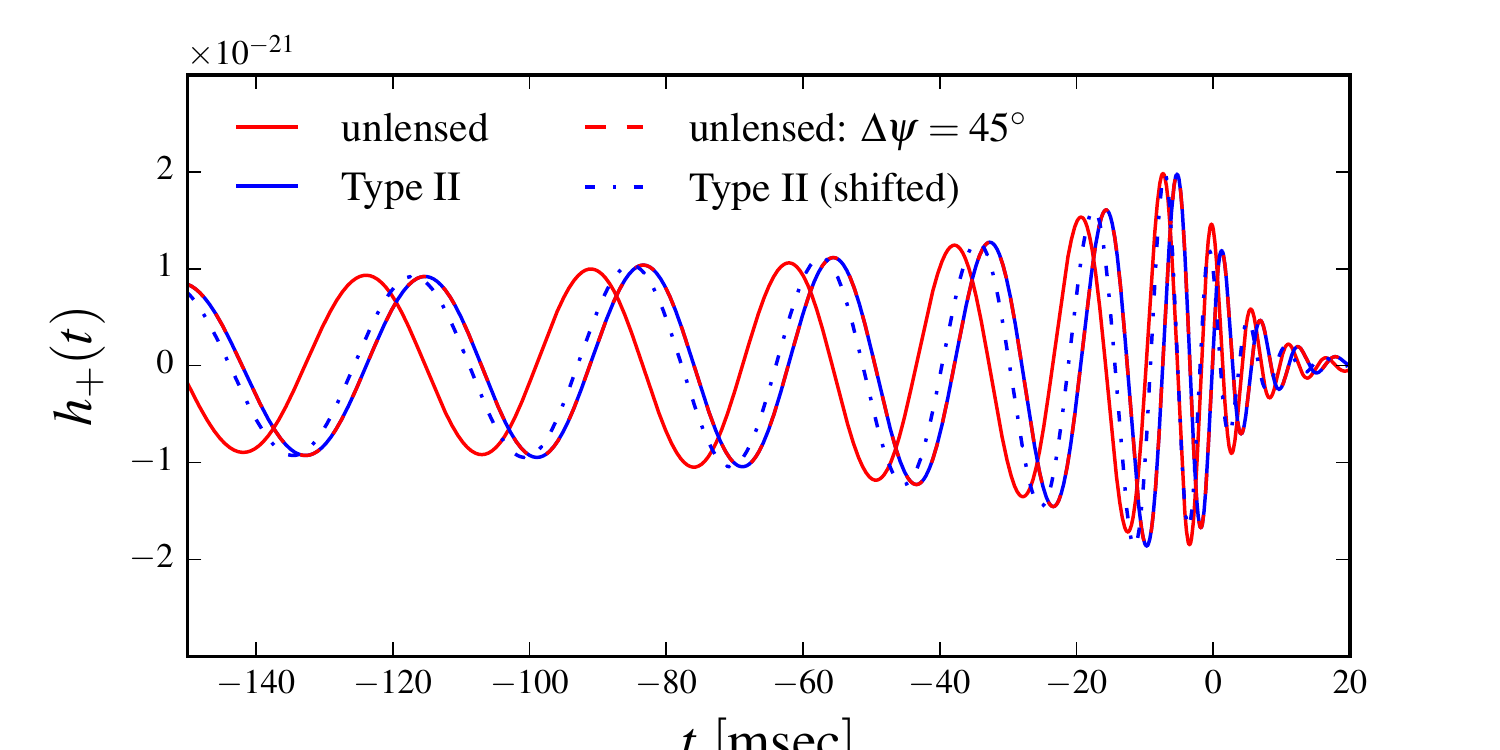}
    \caption{\label{fig:BBH} Merger waveform (``+'' polarization) for a circular non-spinning black-hole binary of $36\,M_\odot$  and $29\,M_\odot$, at $z = 0.5$. The binary is assumed to be at the zenith, with orbital inclination $\iota = 60^\circ$. The waveform is generated from the ${\tt IMRPhenomD}$ model~\cite{Santamaria:2010yb, Kumar:2016dhh}. The image shows the unlensed strain waveform (red),  that for a Type-II image (blue), unlensed but with the azimuthal angle increased by $45^\circ$ (red dashed, on top of blue). Also shown is a shifted Type-II image, to emphasize the distortion in the accumulative phase when the cycle of maximum amplitude is aligned with that of the unlensed waveform (blue dash-dotted). Lensing magnification is not included.}
  \end{center}
\end{figure}
%%%%%%%%%%%%%%%%%%%%

The usual decomposition of the strain signals of the ``$+$" and ``$\times$'' polarizations into spin-weighted spherical harmonics is
\ba
h_+(\bfn, t) \pm i\,h_\times(\bfn, t) = \sum^\infty_{\ell=2}\,\sum^{\ell}_{m=-\ell}\, h^{\pm2}_{\ell m}(t)\,{}_{\mp 2} Y_{\ell m}(\bfn),
\ea
where $\bfn$ is the line-of-sight direction. Since the temporal series $h_+$ and $h_\times$ are real-valued, $[ h^{\pm2}_{\ell m}(t)]^* = (-)^m\, h^{\mp 2}_{\ell,-m}(t)$. Instead of spin-weighted quantities, we can directly work with the transverse trace-free metric perturbation:
\ba
h_{ij}(\bfn, t) & = &  \sum^\infty_{\ell=2}\,\sum^{\ell}_{m=-\ell}\,\left[ \frac{1}{\sqrt 2} \left( h^{2}_{\ell m} + h^{-2}_{\ell m} \right)\,Y^{TE}_{\ell m,ij}(\bfn) \right. \en
&& \left. + \frac{1}{\sqrt 2\,i} \left( h^{2}_{\ell m} - h^{-2}_{\ell m} \right)\,Y^{TB}_{\ell m,ij}(\bfn) \right],
\ea
where $Y^{TE/TB}_{\ell m,ij}(\bfn)$ are transverse trace-free spherical harmonic tensors of electric/magnetic type as defined in Ref.~\cite{Dai:2012bc}. 

GW radiation is often strongly dominated by the electric quadrupole\footnote{Historically, the nomenclature of Ref.~\cite{1963PhRv..131..435P} for the electric/magnetic types is opposite to what we use here.} $TE$ with $\ell = 2$, since the $\ell=2$ magnetic quadrupole and higher-order harmonics $\ell >2$ are suppressed due to non-relativistic source velocities. If we neglect these contributions, we can restrict ourselves to $\ell = 2$ and assume $h^{2}_{2 m} = h^{-2}_{2m}$. Combining this with the reality condition, we have $[h^{2}_{2m}]^* = (-)^m\,h^{2}_{2,-m}$. This can be explicitly verified using the quadrupole formula for GW radiation~\cite{PhysRev.136.B1224}. 

{\it(a) Circular binaries}: For plane-symmetric systems with nearly circular motion, the $m = \pm2$ modes usually dominate. The wave is then determined by only one complex-valued multipole moment $h^2_{22}(t)$,
\ba
\label{eq:h22only}
h_+(\bfn, t) & = & \sqrt{\frac{5}{64\,\pi}}\,2\,\left( 1 + \cos^2\iota \right)\,{\rm Re}\left[ h^2_{22}(t)\,e^{i\,2\,\psi} \right], \en
h_\times(\bfn, t) & = & - \sqrt{\frac{5}{64\,\pi}}\,4\,\cos\,\iota\,{\rm Im}\left[ h^2_{22}(t)\,e^{i\,2\,\psi} \right].
\ea
Here $\iota$ and $\psi$ are the inclination and the azimuthal angles of the line of sight $\bfn$ with respect to the orbital plane.

\refeq{h22only} implies that {\it if the Fourier decomposition of $h^2_{22}(t)$ consists of only positive (negative) frequencies, then multiplication by a factor of $-i$ ($+i$) in Fourier space (such as for a Type-II image) is equivalent to a change of the azimuthal angle $\psi$ by $-\pi/4$ ($+\pi/4$).} In such a case, a Type-II image would not appear unusual, but rather a different source orientation would be inferred (similarly a Type-III image would fit a source with the angle $\psi$ rotated by $\pi/2$). This situation applies to inspirals of circular, non-precessing binaries, the category of astrophysical sources most relevant to observations. This is because under the quadrupole approximation and in a coordinate system where the orbit lies in the $x-y$ plane, $h^2_{22}(t)$ is sourced by a particular combination of the mass quadrupole moments $\ddot Q_{xx} - \ddot Q_{yy} + 2\,i\,\ddot Q_{xy}$. For a circular binary, this combination has a harmonic time dependence $\propto \exp{[2\,i\,\Omega\,t]}$ with $\Omega$ being the orbital (angular) frequency, and thus satisfies the aforementioned condition.

The above conclusion holds to high accuracy even through the merger and ringdown of a circular black-hole binary, regimes in which the weak-field approximation fails and numerical relativity is needed (see \reffig{BBH}). 

%%%%%%%%%%%%%%%%%%%%
\begin{figure}[t]
  \begin{center}
  \hspace{-0.5cm}
    \includegraphics[width=\columnwidth]{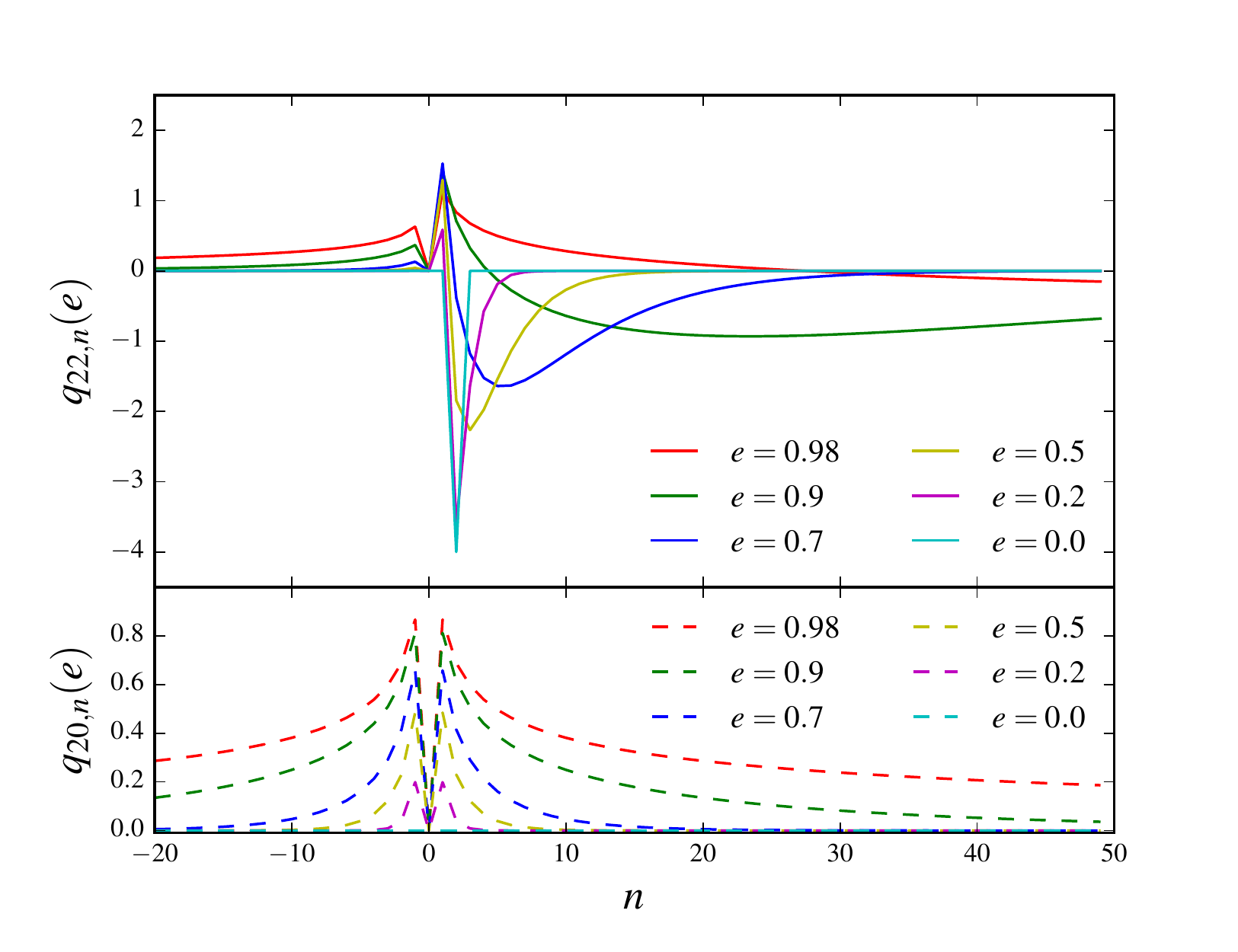}
    \caption{\label{fig:harmonicspec} Fourier amplitudes $q_{22,n}(e)$ and $q_{20,n}(e)$ for the $n$th harmonic sourcing the $m=2$ radiation mode (upper panel) and the $m=0$ radiation mode (lower panel), respectively, as a function of orbital eccentricity $e$.}
  \end{center}
\end{figure}
%%%%%%%%%%%%%%%%%%%%

{\it (b) Eccentric binaries}: The degeneracy between the lensing-induced phase shift and source orientation does not necessarily hold for general GW sources. As a specific example, let us study non-relativistic binaries that can be approximated as two point masses in Keplerian motion (restricted to quadrupole radiation). For nonzero orbital eccentricity $e$, the waveform for a Type-II image is no longer equivalent to that of a binary with a different azimuthal angle $\psi$ for two reasons: (1) the orbital motion is not purely-harmonic in time; (2) the $m=0$ radiation mode turns on in addition to the $m=\pm 2$ modes.

Firstly, consider the mass quadrupole that sources $h^2_{22}$. For a bound orbit $0 \leq e < 1$, it can be expanded in a Fourier series, $\ddot Q_{11} - \ddot Q_{22} + 2\,i\,\ddot Q_{12}  =  \mu^2\,a^2\,\Omega^2\,\sum^{+\infty}_{n = -\infty}\,q_{22,n}(e)\,e^{i\,n\,\Omega\,t} $, where $\mu$ is the reduced mass, $a$ is the semi-major axis, and $q_{22,n}(e)$ is the amplitude for the $n^{\rm th}$ harmonic. For $e \ll 1$, the amplitudes of the $n < 0$ harmonics are negligible compared to those of the $n>0$ harmonics. However, as $e$ approaches unity, the negative-frequency modes become increasingly important (\reffig{harmonicspec}), and the same-sign-frequency condition mentioned before is violated.

Secondly, for eccentric binaries $\ddot Q_{11} + \ddot Q_{22} = \mu^2\,a^2\,\Omega^2\,\sum^{+\infty}_{n = -\infty}\,q_{20,n}(e)\,e^{i\,n\,\Omega\,t}$ is non-vanishing (scales linearly with small $e$), and sources the $m=0$ radiation mode. For large eccentricity $e \lesssim 1$, this mode significantly contributes to the plus polarization in \refeq{h22only}, $\Delta h_+(\bfn, t) = \sqrt{15/32\,\pi}\,\sin^2\iota\,h^2_{20}(t)$, and the Type-II image's distortion to this cannot be compensated by a change in $\psi$.

%%%%%%%%%%%%%%%%%%%%
\begin{figure}[t]
  \begin{center}
  \vspace{-0.2cm}  
  \hspace{-0.5cm}
    \includegraphics[width=\columnwidth]{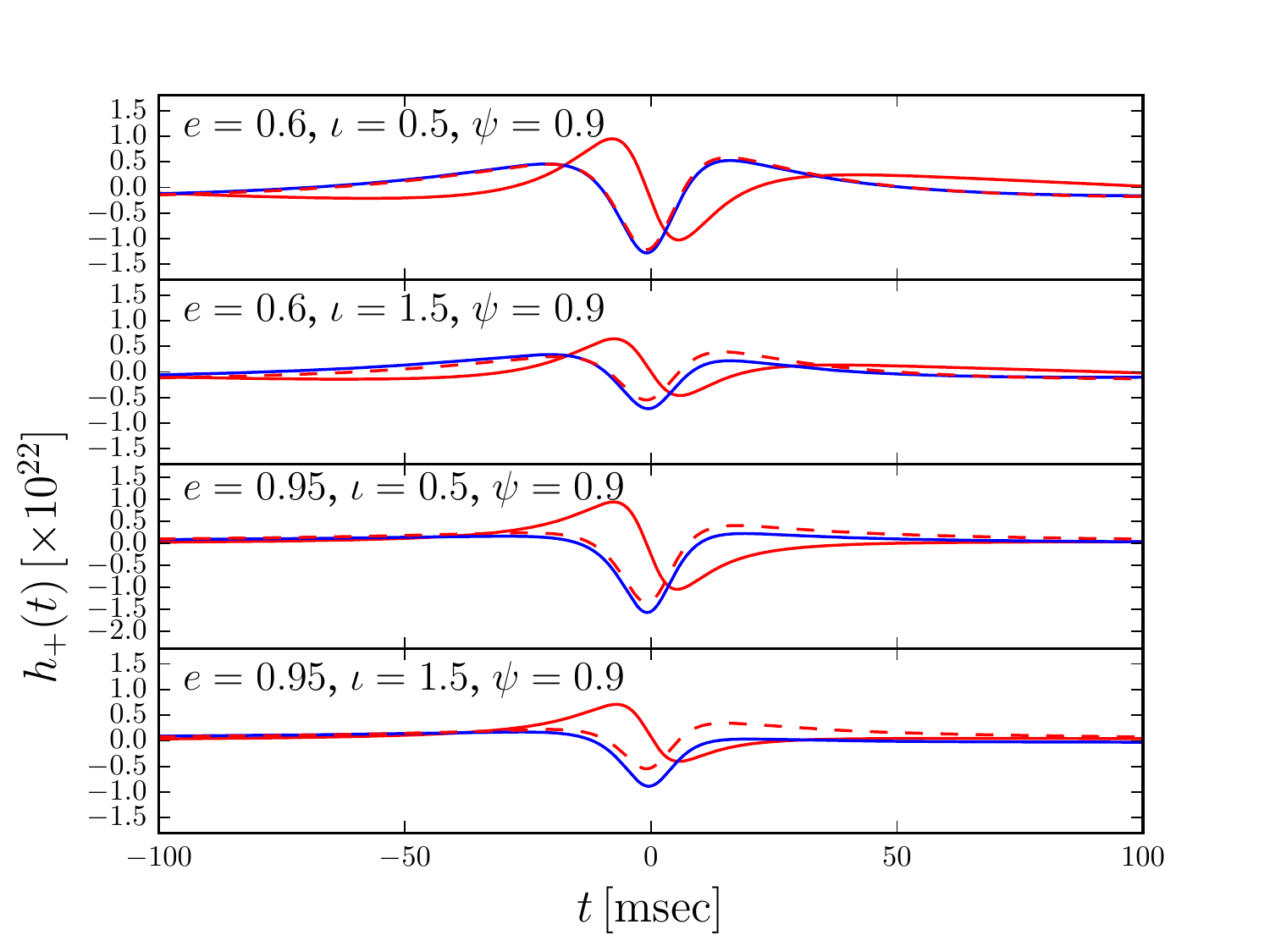}
    \caption{\label{fig:emrb} GW burst produced during the mutual periapsis passage of a pair of $M = 30\,M_\odot$ black holes, separated by $8\,M$  at the pericenter and seen from $z = 1$: unlensed waveform (red), that for a Type-II image (blue), and the unlensed waveform but with $\psi$ rotated by $45^\circ$ (red dashed). We show cases of moderate and large eccentricity ($e=0.6$ and $e=0.95$, respectively) as well as moderate and large inclination ($\iota=0.5$ and $\iota = 1.5$, respectively). As earlier, we neglect lensing magnification, without which a single burst would have matched filtering SNR $\sim 10$ with the Einstein Telescope.}
  \end{center}
\end{figure}
%%%%%%%%%%%%%%%%%%%%

As an application, consider the case of lensed GWs from such eccentric binaries. Such systems can form via few-body interactions in dense stellar environments~\cite{wen2003eccentricity, o2009gravitational, 2013ApJ...777..103T, samsing2014formation, 2016ApJ...816...65A}, and are interesting targets for aLIGO/AdV and its successors. These emit strong GW bursts during every periapsis passage. \reffig{emrb} shows lensed (Type-II) and unlensed burst waveforms for such a system, calculated in the quadrupole approximation \cite{peters1963gravitational}. For moderate eccentricity, the effect of lensing mimics a rotation of the orbit's azimuthal orientation by $\pi/4$. For eccentricity very close to unity, $1-e \ll 1$, the effect of lensing is significantly distinguishable from any rotation of $\psi$. Nearly edge-on orbits ($\iota \approx \pi/2$) show the largest difference from a rotation of $\psi$ by $\pi/4$, due to the $m=0$ radiation mode.  

For unbound orbits $e>1$, the waveform has a non-oscillatory memory~\cite{turner1977gravitational}, for which a treatment based on the short-wavelength limit of the diffraction integral is inapplicable. How lensing distorts a gravitational waveform with memory is left out of the scope of this paper.

{\it Conclusion.}---When GWs from binaries are strongly lensed, it is commonly expected that a parameter inference routine that does not account for lensing would infer consistent sky localizations\footnote{Even for futuristic detector networks, the angular resolution is insufficient to resolve typical image separations.} and best-fit dimensionless parameters\footnote{Inferred dimensionful parameters will necessarily be different due to different magnification factors $\mu_j$ for individual images~\cite{Dai:2016igl}.} for the multiple images. We have shown that when the sources are circular binaries, the inferred azimuthal angle $\psi$ for Type-II (-III) images will be offset by $\pi/4$ ($\pi/2$) relative to that of the earliest image. This will allow us to determine the topological type of each image; together with time delays and relative magnifications, this will provide valuable information about the lens model. In the case of multiply-imaged GWs from eccentric systems (and even for precessing binaries), Type-II images are deformed in a characteristic manner that cannot be reproduced by a simple change of the source parameters. Measurements of these lensing distortions to the waveforms of Type-II images would verify that gravitational waves, as linear and non-dispersive metric waves in the weak-field regime, propagate through background space-time as predicted by general relativity, in an identical manner to scalar or electromagnetic waves.

%%%%%%%%%%%%%%%%%%%%%%%%%%%%%%%%%%%%%%%%%%%%%%%%%%%%%%%%%%%%%%%%%%%%%%%%%%%%%%%%%%%

\begin{acknowledgments}

{\it Acknowledgment.}---We thank Aaron Zimmerman for reading this draft and offering valuable comments. LD is supported by the NASA Einstein Postdoctoral Fellowship under grant number PF5-160135 awarded by the Chandra X-ray Center, which is operated by the Smithsonian Astrophysical Observatory for NASA under contract NAS8-03060. TV acknowledges support from the Schmidt Fellowship and the Fund for Memberships in Natural Sciences at the Institute for Advanced Study.

\end{acknowledgments}

%%%%%%%%%%%%%%%%%%%%%%%%%%%%%%%%%%%%%%%%%%%%%%%%%%%%%%%%%%%%%%%%%%%%%%%%%%%%%%%%%%

%------------------------------------------------------------------------------
\bibliographystyle{apsrev4-1-etal}
\bibliography{reference_saddle}
%------------------------------------------------------------------------------

\end{document}